# English Accent Accuracy Analysis in a State-of-the-Art Automatic Speech Recognition System

Guillermo Cámbara[1,2], Alex Peiró-Lilja[1], Mireia Farrús[3], Jordi Luque[2,4]

[1]*Universitat Pompeu Fabra,* [2]*Telefónica Research,* [3]*Universitat de Barcelona,*
[4]*Universitat Politècnica de Catalunya*

Nowadays, research in speech technologies has gotten a lot out thanks to recently created public domain corpora that contain thousands of recording hours. These large amounts of data are very helpful for training the new complex models based on deep learning technologies. However, the lack of dialectal diversity in a corpus is known to cause performance biases in speech systems [5], mainly for underrepresented dialects, for example [6] or [7].

In this work, we propose to evaluate a state-of-the-art automatic speech recognition (ASR) deep learning-based model, using unseen data from a corpus with a wide variety of labeled English accents from different countries around the world. The model has been trained with 44.5K hours of English speech from an open access corpus called Multilingual LibriSpeech [4] (MLS), showing remarkable results in popular benchmarks. We test the accuracy of such ASR against samples extracted from another public corpus that is continuously growing, the Common Voice [2] dataset (CV). Then, we present graphically the accuracy in terms of Word Error Rate (WER) of each of the different English included accents, showing that there is indeed an accuracy bias in terms of accentual variety, favoring the accents most prevalent in the training corpus.

The MLS corpus [4] is a larger version of the LibriSpeech dataset [1] derived from the LibriVox audio books, this time with transcribed recordings of 8 languages. As suggested in the LibriVox wiki[1], there might be a higher representation of accents from America (48 speakers between US and Canada) and the UK (26 speakers), than from other countries like India or Philippines (1-2 speakers). As opposed to the MLS dataset, the CV corpus used for evaluation does have accent labels, at least in 321K samples out of its total. The data consist of around 900 hours of recorded speech from volunteers reading scripted texts.

The ASR model pretrained with the MLS dataset (see architecture details in [4]) is used for transcribing the audio utterances in the CV corpus. To continue with, the WER is used to assess the accuracy of the neural ASR. This metric is the summation of word substitutions (S), deletions (D) and insertions (I) in the ASR transcript, divided by the total number of words in the ground truth transcript (N):

$$WER = \frac{S+D+I}{N}$$

WER is computed on every test utterance, and the mean of every accent group is shown in Figure 1.

As hypothesized, the best WER scores are achieved by the most likely represented accents in the training corpora, namely the American accents (US and Canada), plus accents from Ireland, England and New Zealand. Moreover, we see that the WER can degrade up to an absolute 10% for accents with phonetic and prosodic characteristics further from American and UK English, like Asian accents. This highlights that still newer models present accent bias, a problem that could be better tackled using large corpora with rich metadata like the CV dataset. This inspires further research in finding which phonetic and prosodic traits are most impactful for speech recognition performance, and exploring which methods could be more efficient for solving such biases, besides straight data collection.

---

[1] https://wiki.librivox.org/index.php/Accents_Table

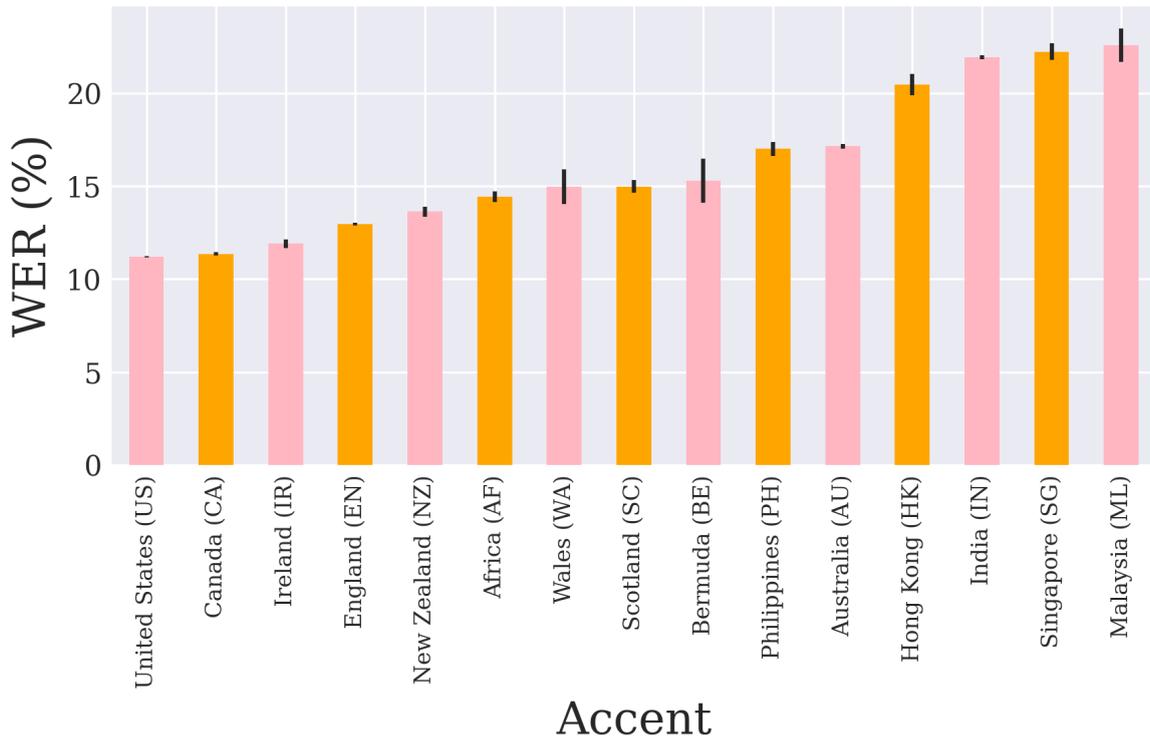

Figure 1. Word Error Rate (%) obtained from the tested ASR for every English accent found in the Common Voice corpus. Standard deviation of the mean is represented as error bars for every accent class.

| Accent | US | CA | IR | EN | NZ | AF | WA | SC | BE | PH | AU | HK | IN | SG | ML |
|---|---|---|---|---|---|---|---|---|---|---|---|---|---|---|---|
| # utt | 171K | 21K | 4K | 47K | 3K | 3K | 0.3K | 2K | 0.2K | 2K | 24K | 1K | 34K | 2K | 0.6K |
| # spkr | 3747 | 445 | 92 | 1079 | 75 | 109 | 33 | 75 | 22 | 59 | 305 | 72 | 1068 | 33 | 46 |

Table 1. Number of sampled utterances and speakers from the Common Voice corpus, for each accent group, being the first counted by thousands.